\documentclass[a4,useAMS,usenatbib]{mn2e}

\usepackage[english]{babel} \usepackage{subfigure}
\usepackage{graphicx}

\usepackage[fleqn]{amsmath} 
\usepackage{color}

\usepackage[varg]{txfonts}

\newcommand{\aj}{AJ} 
\newcommand{\aap}{A\&A} 
\newcommand{\apj}{ApJ} 
\newcommand{\apjl}{ApJL} 
\newcommand{\mnras}{MNRAS} 
\newcommand{\nat}{Nature} 

\newcommand{\noi}{}
\newcommand{\vect}{\bmath}

\newcommand{\msun}{M_\odot}

\newcommand{\kms}{\, {\rm km\, s}^{-1}}
\newcommand{\h}{\,h_{70}}
\newcommand{\hmm}{\h^{-1}}
\newcommand{\Mpc}{\, {\rm Mpc}}

\title[Detection of a Dark Substructure through Gravitational Imaging]{Detection of a Dark Substructure through Gravitational Imaging}

\author[S. Vegetti, L.V.E. Koopmans, A. Bolton, T. Treu, R. Gavazzi]{
S. Vegetti$^{1}$\thanks{E-mail: vegetti@astro.rug.nl}, L.V.E. Koopmans$^{1}$, A. Bolton$^{2}$,
T. Treu$^{3}$ \& R. Gavazzi$^{4}$\\
$^{1}$Kapteyn Astronomical Institute, University of Groningen, P.O. Box 800, 9700\,AV Groningen, the Netherlands\\
$^{2}$Institute for Astronomy, University of Hawaii, 2680 Woodlawn Drive, Honolulu, HI 96822-1897, USA\\
$^{3}$Department of Physics, University of California, Santa Barbara, CA 93101, USA\\
$^{4}$Institut d'Astrophysique de Paris, CNRS, UMR 7095, Universit«e Pierre et Marie Curie, 98bis Bd Arago, 75014 Paris, France}
  
\begin{document}

\date{Accepted  ... Received ...; in original form ....}

\pagerange{\pageref{firstpage}--\pageref{lastpage}} \pubyear{2002}

\maketitle

\label{firstpage}

\begin{abstract}
We report the detection of a dark substructure -- undetected in the HST-ACS F814W image -- in the gravitational lens galaxy SDSSJ0946+1006 (the ``Double Einstein Ring''), through direct gravitational imaging. The lens galaxy is of particular interest because of its relative high inferred fraction of dark matter inside the effective radius. 
The detection is based on a Bayesian grid reconstruction of the two-dimensional surface density of the galaxy inside an annulus around its Einstein radius.  
The detection of a small mass concentration in the surface density maps has a strong statistical significance. We confirm this detection by modeling the substructure with a tidally truncated pseudo-Jaffe density profile; in that case the substructure mass is $M_{\rm{sub}}=(3.51\pm 0.15)\times 10^9\msun$, located at $(-0.651\pm0.038,1.040\pm0.034)$'', precisely where also the surface density map shows a strong convergence peak (Bayes factor $\Delta\log{\cal{E}}= -128.0$; equivalent to a $\sim$16--$\sigma$ detection). We set a lower limit of ${\rm (M/L)}_{{\rm V},\odot}\ga 120~ \msun/{\rm L}_{{\rm V},\odot}$ (3--$\sigma$) inside a sphere of 0.3 kpc  centred on the substructure ($r_{\rm tidal}$=1.1\,kpc). The result is robust under substantial changes in the model and the data-set (e.g.\ PSF, pixel number and scale, source and potential regularization, rotations and galaxy subtraction). It can therefore not be attributed to obvious systematic effects. Our detection implies a dark matter mass fraction at the radius of the inner Einstein ring of $f_{\rm CDM}=2.15^{+2.05}_{-1.25}$ percent (68\% C.L) in the 
mass range $4\times10^6\msun$ to $4\times10^9\msun$ assuming $\alpha=1.9\pm0.1$ (with $dN/dm\propto m^{-\alpha}$). Assuming a flat
prior on $\alpha$, between 1.0 and 3.0, increases this to $f_{\rm CDM}=2.56^{+3.26}_{-1.50}$ percent (68\% C.L). The likelihood ratio
is 0.51 between our best value ($f_{\rm CDM} = 0.0215$) and that from simulations ($f_{\rm sim} \approx 0.003$). Hence the inferred mass fraction, admittedly based on a single lens system, is large but still consistent with predictions. We expect to further tighten the substructure mass function (both fraction and slope), using the large
number of systems found by SLACS and other surveys.

\end{abstract}


\section{Introduction}

In the process of building a coherent picture of galaxy formation and evolution, early-type galaxies play a crucial role. Often unfairly referred to as \emph{red and dead} objects, many aspects about their structure and formation are still unknown. What is the  origin of the tight empirical relations between their global properties \citep{Djorgovski87,Dressler87,Magorrian98,Ferrarese00,Gebhardt00,Bower92,Guzman92,Bender93}?  How do massive early-type assemble? What is the fraction of mass substructure populating the haloes of early-type galaxies and is this in agreement with the CDM paradigm \citep{Kauffmann93,Moore99,Klypin99,Moore01,Maccio06,Diemand08,Springel08,Xu09}?

\noi Gravitational lensing, especially in combination with other techniques, provides an invaluable and sometimes unique insight in answering these questions \citep[e.g.][and references therein]{Rusin05,Treu04,Koopmans09}.

\noi At the level of small mass structure lensing stands out as a unique investigative method; different aspects of the lensed images can be analysed to extract information about the  clumpy component of galactic haloes. 
Flux ratio anomalies, astrometric perturbations and time-delays, in multiple images of lensed quasars, can all be related to substructure at scales smaller than the images separation \citep{Mao98,Bradac02, Chiba02, Dalal02, Metcalf02, Keeton03, Kochanek04, Bradac04, Keeton05,McKean07,Chen07,2009MNRAS.394..174M}.

\noi As described by \citet{Koopmans05} and \citet{Vegetti09a}, the information contained in multiple images and Einstein rings of extended sources, can also be used. While the former three approaches only provide a statistical measure of the lens clumpiness, the latter allows one to identify and quantify of each single substructure, measuring for each of them the mass and the position on the lens plane. Both approaches are however complementary in that the former is more sensitive to low-mass perturbations, which are potentially present in large numbers, whereas the latter is sensitive to the rarer larger scale perturbations.

\noi The method of direct gravitational imaging of the lens potential -- shortly described in the following section -- represents an objective approach to detect dark and luminous substructures in individual lens systems and allows on to statistically constrain the fraction of galactic satellites in early-type galaxies. Extensively described and tested in \citet{Vegetti09a}, the procedure is here applied to the study of the double ring SDSSJ0946+1006 \citep{Gavazzi08} from the sample of the \emph{Sloan Lens ACS Survey }(SLACS), yielding the first detection of a dwarf satellite through its gravitational effect only, beyond the Local Universe.

\noi The layout of the paper is as follows: in Section 2 we provide a general description of the method. In Section 3 we introduce the analysed data and in Section 4 we discuss the results of the modelling under the assumption of a smooth lens potential.  In Section 5 we describe the detection of the substructure. In Section 6 we present the error analysis and the model ranking. In Section 7 we discuss implication for the CDM paradigm and in Section 8 we conclude. Through out the paper we assume the following cosmological parameters with $H_0 = 73\,\kms\Mpc^{-1}$, $\Omega_{\rm m}=0.25$ and $\Omega_\Lambda=0.75$. 

\section{The method}

In this section we provide a short introduction to the lens modelling method. 
The main idea behind the method of ``gravitational imaging'' is that effects related to the presence of dwarf satellites and/or CDM substructures in a lens galaxy can be modelled as local perturbations of the lens potential and that the total potential can be described as the sum of a smooth parametric component with linear corrections defined on a grid. 
We refer to \citet{Vegetti09a} for a more complete discussion.

\subsection{Source and potential reconstruction}\label{sec:reconst}

As shown in \citet{2001ASPC..237...65B}, \citet{Koopmans05}, \citet{Suyu206} and \citet{Vegetti09a}, it is possible to express the relation between perturbations in the lensed data ($\delta d$; i.e. perturbations of the surface brightness distribution of the lensed images), the unknown source surface brightness distribution ($s$) and perturbations in the lens potential ($\delta \psi$) as a set of linear equations $\delta d = -\nabla s \cdot \nabla \delta \psi$. Through the Poisson equation $\delta \psi$ can be turned into a relation with the convergence correction $\delta \kappa = \nabla^{2} \delta \psi /2$.

For a fixed form of the lensing potential and regularization, the
inversion of these equations leads to the simultaneous reconstruction
of the source and a potential correction.  The source grid is defined
by a Delaunay tessellation which automatically concentrates the
computational effort in high magnification regions while keeping the
number of degrees of freedom constant, which is critical in assessing
the Bayesian posterior probability and evidence for the model
\citep[see][]{Vegetti09a}.  The procedure is embedded in the framework
of Bayesian statistics which allows us to determine the best set of
non-linear parameters for a given potential and the linear parameters
of the source, to objectively set the level of regularization and to
compare different model families \citep{MacKay92, MacKay03,
Suyu206,Brewer06}. Specifically, for a particular lens system we wish
to objectively assess whether it can be reproduced with a smooth
potential or whether mass structure on smaller scales has to be
included in the model.

\noi The modelling is performed via a four steps procedure: (i) We start by choosing a form for the parametric smooth lens density profile, generally an elliptical power-law, and we determine the non-linear parameters and level of source regularisation that maximize the Bayesian evidence, through a non-linear optimization scheme. (ii) In the case that this model is too simple and significant image residuals are left, we allow for grid-based potential corrections. This leads to the initial detection and localization of possible substructures. (iii) The substructure masses and positions are then more precisely quantified by assuming a tidally truncated pseudo-Jaffe (PJ) profile \citep{Dalal02} and by simultaneously optimising for the main lens galaxy and substructure parameters, i.e. its mass $M_{\rm{sub}}$ and position on the lens plane $(x_{\rm{sub}};y_{\rm{sub}})$. (iv) Finally the two models, i.e. the single power-law (PL) and the power-law plus PJ substructures (PL+PJ), are compared through their total marginalized Bayesian evidences ($\cal E$), that represent the (conditional) probabilities of the data marginalized over all variable model parameters. 

\subsection{Detection Threshold of Mass Substructure}

The method has a mass detection threshold to substructure that depends on the signal-to-noise ratio and spatial resolution of the lensed images; for typical HST (e.g. SLACS) data quality the mass detection threshold for a substructure located on the Einstein ring and with a pseudo-Jaffe density profile is of the order of a few times $10^8\msun$ and quickly increases with the distance from the lensed images \citep[see][]{Vegetti09a} because
of the decrease in the image surface brightness and local magnification.

\noi Despite having been developed with the specific task of identifying and constraining the fraction of substructure in lens galaxies, this technique can also be used to model complex lens potentials, that are relatively smooth, but do not have the simple symmetries that are often assumed in mass models (e.g.\ elliptical power-law density profiles). As shown in \citet{Barnabe09}, we can also reconstruct the lensed images and the relative sources down to the noise level, even for systems that are highly asymmetric and strongly depart from a power-law density profile. The grid-based potential correction is able to correct the inexact initial choice of the lens potential model and recover existing asymmetries in the mass distribution. 

In the rest of this paper, we use this method to analyse the double Einstein ring system SLACS SDSSJ0946+1006 and search for
deviations from a smooth power-law elliptical mass model.

\section{The data}

In this section we present a brief overview of the double Einstein Ring lens system SLACS SDSSJ0946+1006. We refer to \citet{Gavazzi08} for a more detailed description. 

SLACS selects gravitational lens candidates from the Sloan Digital Sky Survey spectroscopic database on the basis of multiple emission lines in the spectrum at redshifts larger than that of the lower redshift target galaxies \citep{2006ApJ...638..703B}. The system was selected by the presence of multiple emission lines at $z_{s1}=0.609$ in the spectrum of a lensing galaxy at $z_l=0.222$. Subsequently confirmed as a strong lens with ACS on board the HST, the system shows a very peculiar structure in the lensed images which are composed by two concentric partial rings, at radii of $1.43'' \pm 0.01''$ and $2.07''\pm 0.02''$, respectively, from the centre of the lens galaxy. This particular configuration is related to the presence of two sources at different redshifts which are being lensed by the same foreground galaxy (see \citet{Gavazzi08} for the a priori probability of this event in a survey such as SLACS);  the nearest source is lensed into the inner ring (Ring 1), while the second one, further away along the optical axis, is lensed into the outermost ring (Ring 2). Ring 1 with a F814W magnitude $m_1 = 19.784 \pm 0.006$ is one of the brightest lensed sources in the SLACS sample, while Ring 2 with $m_{2} = 23.68 \pm 0.09$ is 36 times fainter. Ring 2 is not observed in the SDSS spectrum, and an upper limit to its redshift $z_{s2}  < 6.9$ was set on the basis of ACS imaging. As inferred by \citet{Gavazzi08} the lens galaxy has a projected dark matter mass fraction inside the effective radius that is about twice the average value of the SLACS lenses \citep{Gavazzi07, Koopmans06}, i.e. $f_{\rm{DM}}\left(< R_{\rm eff}\right)\approx 73\%\pm9\% $, corresponding to a project dark matter mass approximately equal to $M_{\rm{DM}}\left(< R_{\rm eff}\right)\approx 3.58\times 10^{11} \hmm \msun$. 

The high dark-matter mass fraction makes this system particularly interesting for CDM substructure studies. If the framework of galaxy formation given by N-body simulations is correct (substructure mass function slope $\alpha=1.90$ and projected dark matter fraction in substructure $f=0.3\%$), we would expect, within an annulus of 0.6$''$ centered on the Einstein radius, on average $\mu=6.46\pm0.95$ substructures \citep{Vegetti09b} with masses between $4\times10^6\msun$ and $4\times10^9\msun$ \citep{Diemand07a,Diemand07b,Diemand08}.
Whereas these have typical masses a few times that of the lower limit
on this range, the probability of finding a mass substructure above
$\ga 10^{8} \msun$ is certainly non-negligible. 

\begin{figure}
   \begin{center} 
      \includegraphics[width=1\hsize]{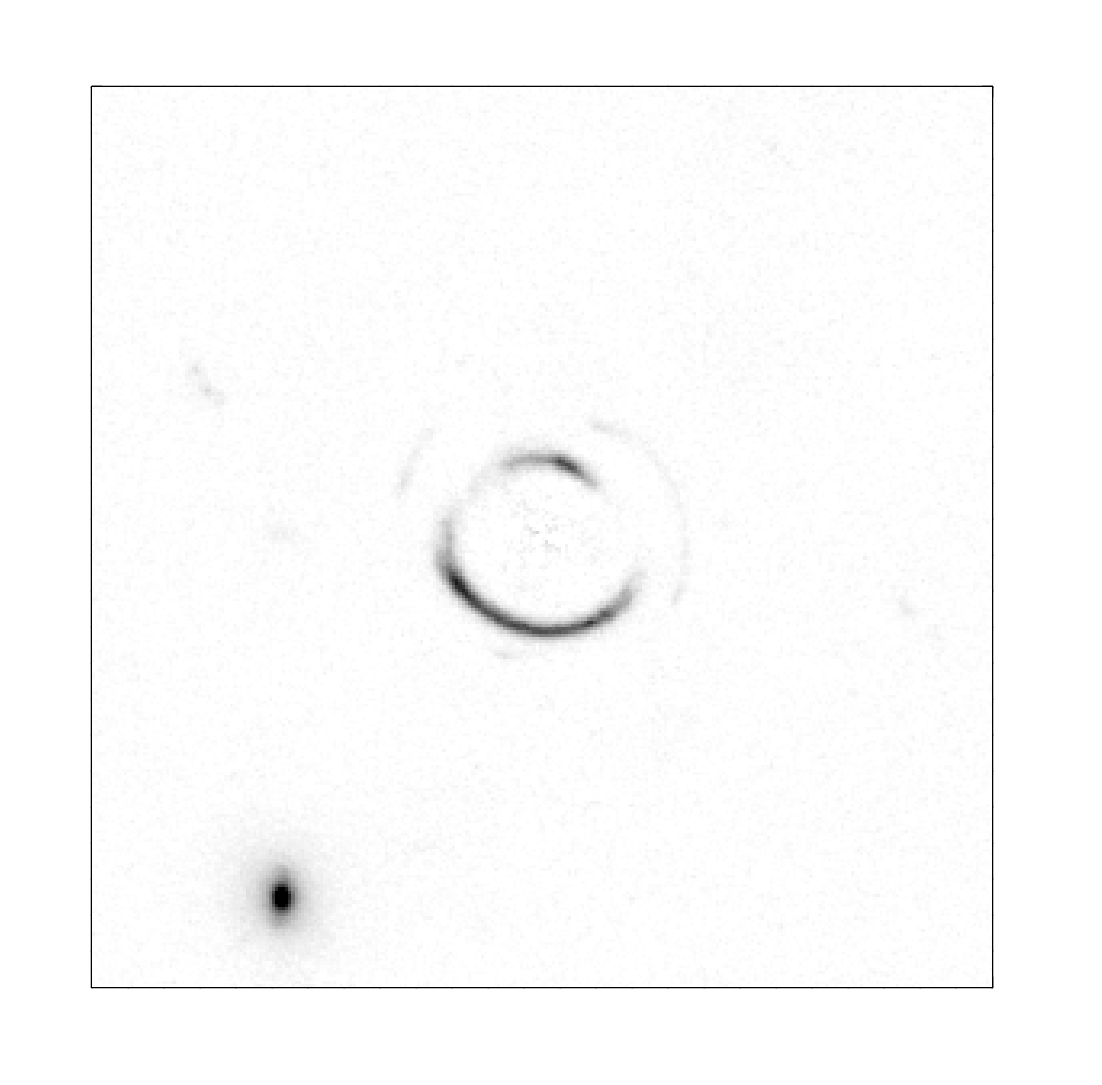}
         \caption{The image of the lens system SDSSJ0946+1006, obtained with HST-ACS through the filter F814W, after subtraction of the lens surface brightness distribution.}
      \label{fig:overview} 
    \end{center}     
 \end{figure}	

\section{Smooth Mass Models}

In this section we describe the details and the results of our analysis. Because of the very low surface brightness, and of the low signal-to-noise ratio of the images associated with Ring2, we limit our study to a tight annulus around Ring 1, in which Ring 2 has been fully excised. This does not affect the lens potential reconstruction which is almost solely constrained by the detailed information given by the high surface brightness distribution of Ring1. Our potential reconstruction
therefore {\sl only} probes the region around the inner ring. The choice
of reconstructing the potential ($\psi$) inside a limited field of view (e.g. mask), rather than the surface mass density ($\kappa$), ensures that the potential and the resulting convergence reconstructions are both unbiased \citep[see][for a detailed discussion about this subtle point]{Koopmans05}.

\subsection{Image Reconstruction}\label{sec:smooth}

At the first level of reconstruction all potential corrections are kept at zero. 
We start by assuming that to first order the lens is well approximated  by a simple smooth elliptical power-law density profile \citep{Barkana98} with a convergence (surface density in terms of critical density $\Sigma_c$) 
\begin{equation}
k(r) = \frac{\alpha}{2\sqrt f r^{2q+1}},
\end{equation}
where $r=\sqrt{x^{2} + y^{2}/f^{2}}$.
The non-linear parameters describing the lens are: the lens strength 
$\alpha$, the position angle $\theta$, the flattening $f$, the centre coordinates $\vect{x_0}$, the projected density slope $q$, the shear strength $\Gamma_{sh}$ and the shear angle $\theta_{sh}$. We do not optimise for the mass centroid, but center on the peak of the surface brightness distribution, as precisely determined from the HST image. 
We show in Section \ref{sec:systematics} that this assumption does not alter the main results of the paper, but reduces our substantial computational load. 

As described in more detail in \citet{Vegetti09a}, the source grid is constructed from a (sub) sample of pixels in the image plane which are cast back to the source plane using the lens equation. The number of grid-points can be objectively set by comparing their Bayesian evidence. In this particular case, we find that using all the image points (e.g. $81\times81$ pixels) is the most appropriate choice. On the image plane the pixel-scale is constant and equal to 0.05''/pixel, while on the source plane the Delaunay triangle-scale is adaptive and depends on the local lensing magnification. We adopt an adaptive curvature regularization, weighting the regularization penalty by the inverse of the image signal-to-noise ratio. We find that this significantly improves the modeling of sharp high dynamic range features in the lensed images, where in general all other forms of regularization (e.g. gradient or unweighted curvature) falter and give much lower evidence values. 

\begin{figure*}
   \begin{center} 
      \includegraphics[width=0.85\hsize]{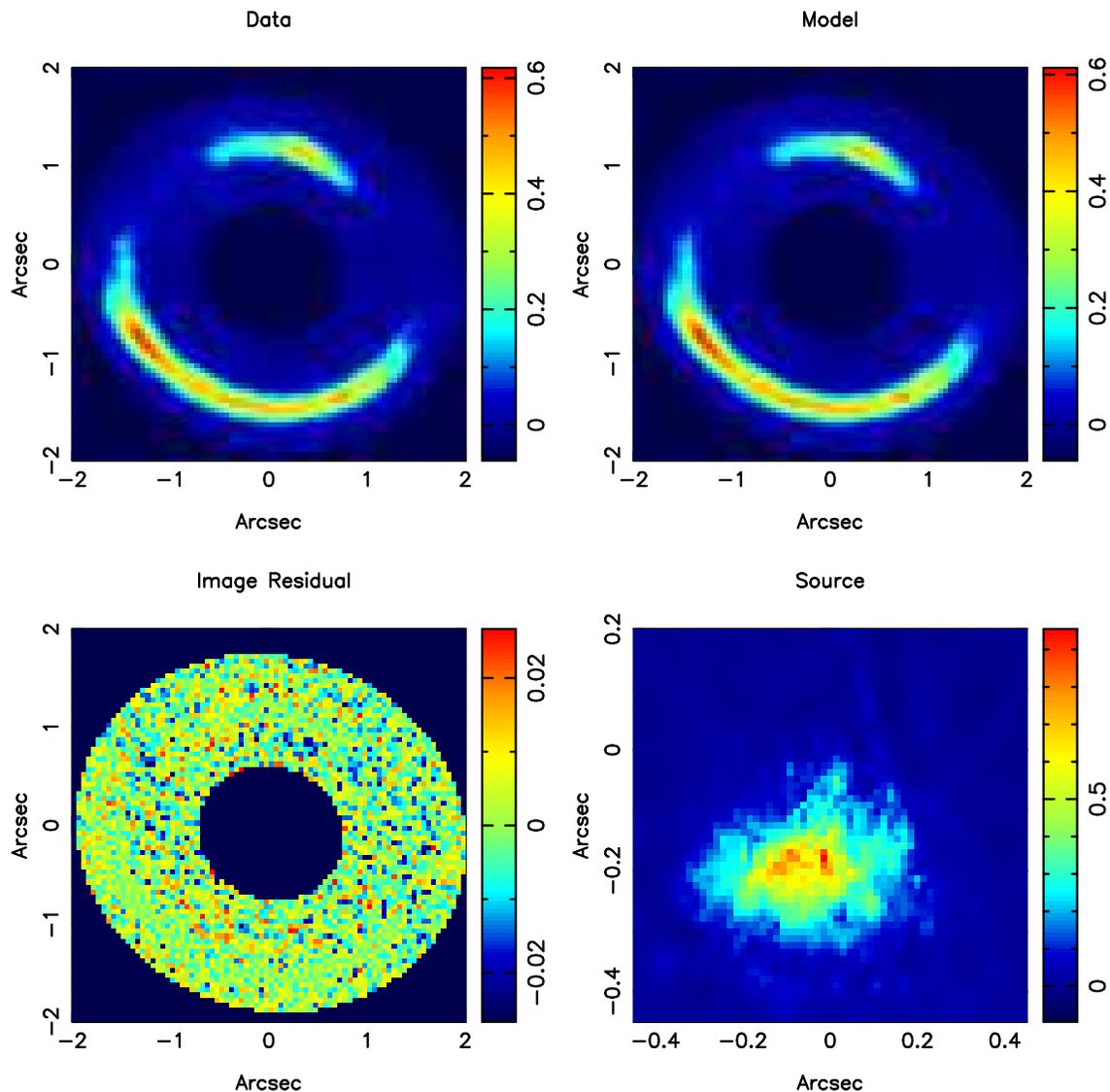}
      \caption{Results of the lens and source reconstruction under the hypothesis of a  
      	smooth potential. The top-right panel shows the original lens data, while the top-left one shows the final
	reconstruction. On the second row 
	the image residuals (left) and the source reconstruction (right) are shown.}
      \label{fig:best_smooth} 
    \end{center}     
 \end{figure*}

We use the results obtained by \citet{Gavazzi08}, for a single lens
plane, as starting point $\vect{\eta_0}$ and then optimize for the
potential parameters and the level of the source regularization. The
resulting source and image reconstruction are presented in
Fig. \ref{fig:best_smooth}.  In Table 1 the recovered lens parameters
and level of source regularization $\left\{ \vect{\eta}_{\rm
b},\lambda_{s,\rm{b}} \right\}$ are listed. The recovered parameters
for the smooth mass component of the lens potential are somewhat
different from the results in \citet{Gavazzi08}, which we attribute to
the fact that \citet{Gavazzi08} makes us of both Einstein rings and
matches conjugate points instead of the full surface brightness
distribution. Some notable results for the smooth mass model are a
density slope $\gamma'\equiv (2 q +1) \approx 2.20$ and a mass axial
ratio of $f=0.96$, indicating that the galaxy is very close to an
isothermal sphere mass model, although has a slightly steeper density
profile. The quantity cannot be compared direclty with the slope
measured by \citet{Gavazzi08} because we are measuring the slope at
the location of the inner ring, while \citet{Gavazzi08} is measuring
the average slope in between the rings.

\subsection{Image Residuals after Reconstruction}\label{sec:resid}

\noi In \ref{fig:best_smooth}, we clearly see remaining image residuals
above the noise level, in particular near the upper-most arc feature. 
The source appears to be a normal, although not completely symmetric,
galaxy. Structure in the source (e.g. brightness peaks and the faint tail-like feature to the upper-right of the source) can also be one-to-one related to structure in the arcs. This provides strong confidence the the overall reconstruction of the system, as being remarkably accurate despite its complexity. The source still shows significant structure on small scales, which is due to a preferred low level of regularisation, when optimizing
for the Bayesian evidence (note that at this level the evidence is simply 
the posterior probability of the free parameters, including the source regularisation).

The image residuals can be related either to different aspects of the reconstruction procedure, for example to the modelling of the PSF, the choice of the simply-parametrized  model for the lens potential, the number and scale of the image pixels, the lens-galaxy subtraction or features in the galaxy brightness profile. To test whether these residuals are related to the presence of substructures, however, we now first proceed by consider a more general model in which we allow for very general potential corrections (see above). We discuss the effects of
systematic errors in a later Section, but stress that non of the above systematic errors are expected to mimic localized lensing features.

\section{The Detection of Mass Substructure}\label{sec:substructure}

From the ``Occam's razor'' point of view, it is more probable that uncorrelated structures in the lensed images are related to local small-scale perturbations in the lens potential, rather than features in the source distribution itself \citep{Koopmans05,Vegetti09a}. It is, therefore, possible to describe galaxy substructure or satellites as linear local perturbations to an overall smooth parametric potential and separate them from changes in the surface brightness distribution due to the source model \citep{Koopmans05}. Given that the remaining image residuals are small, we can assume that  the values for the lens parameters recovered in the previous section are sufficiently close to the real smooth component of the lens potential such that our linearized reconstruction of the source and the grid-based lens-potential corrections are fully justified, as discussed in Section \ref{sec:reconst}.  

\subsection{Grid-based Substructure Modeling}

\begin{figure*}
\begin{center}
\includegraphics[scale=0.6]{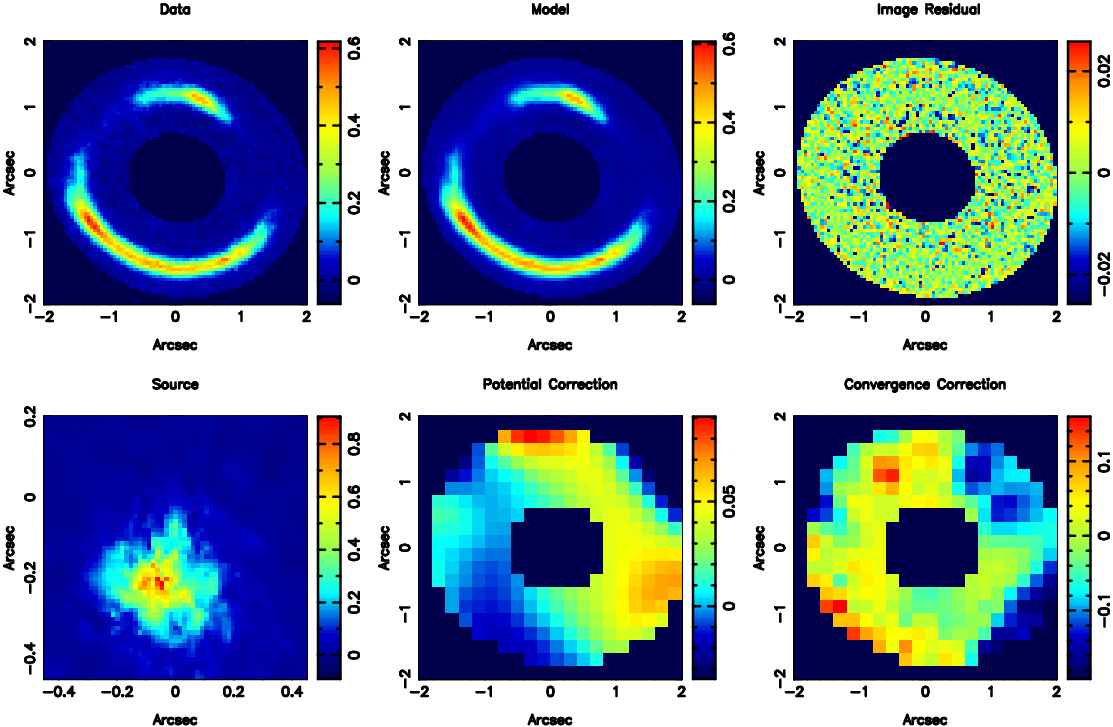}
\includegraphics[scale=0.6]{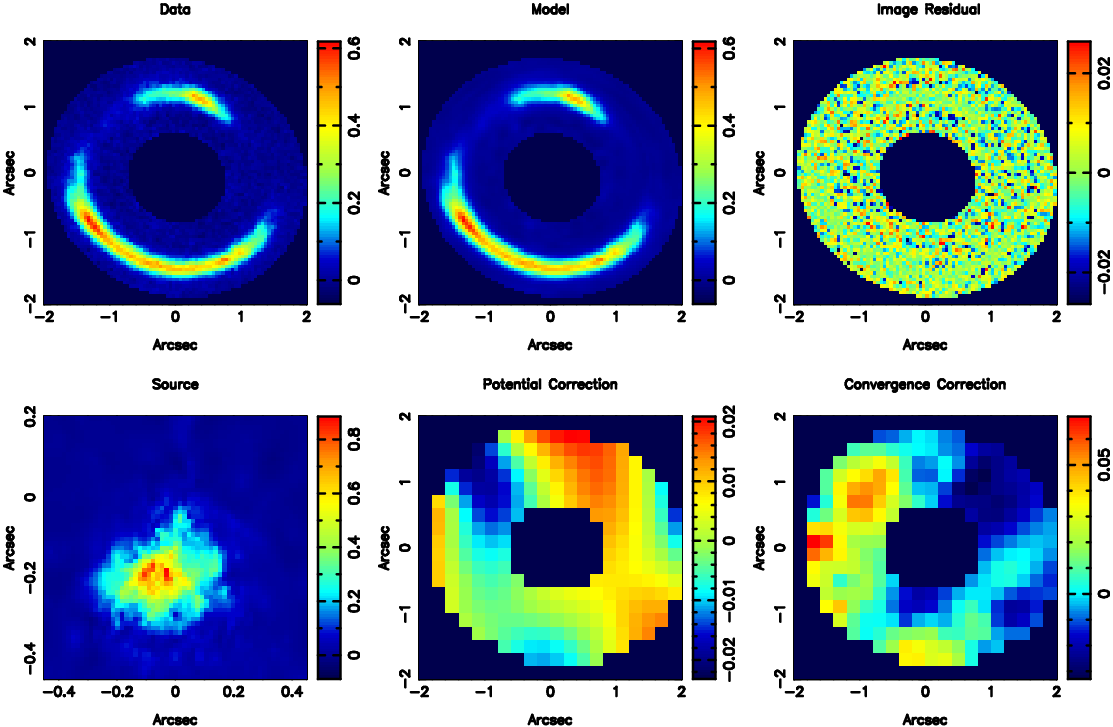}
\includegraphics[scale=0.6]{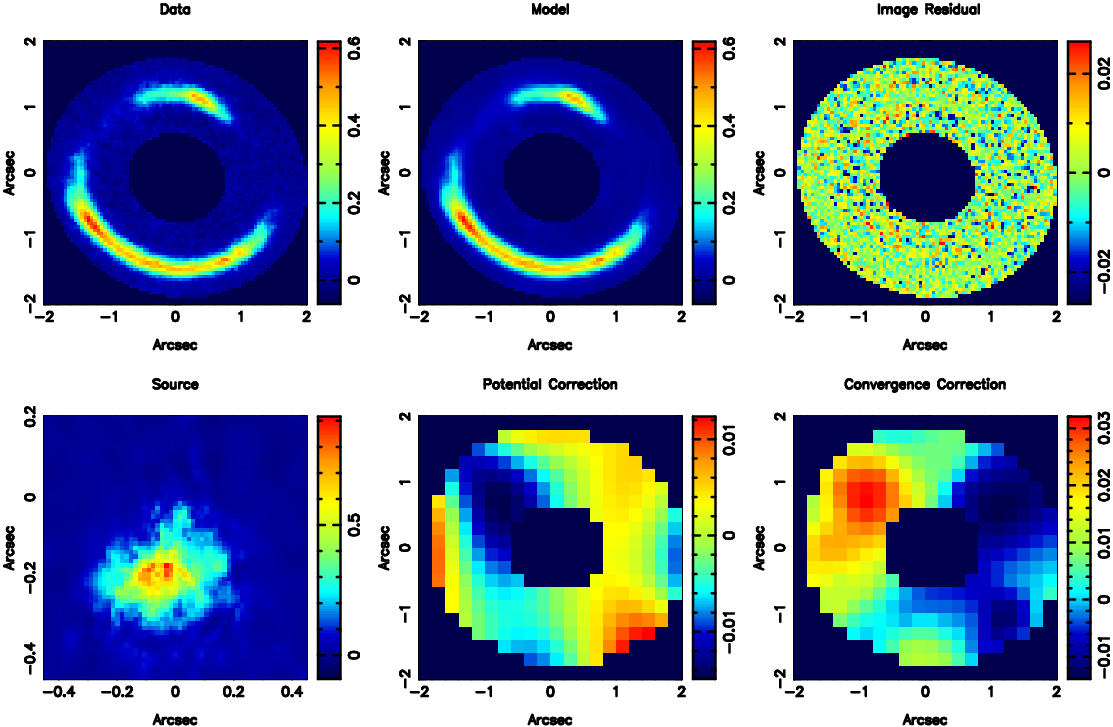}
\end{center}
\caption{Results of the pixelized reconstruction of the source and lens potential corrections for three different value of the potential corrections regularization $\lambda_{\delta\psi}=10^7$ (top left panels), $\lambda_{\delta\psi}=10^8$ (top right panels) and $\lambda_{\delta\psi}=10^9$  (low panels). The top-right panel shows the original lens data, the middle one shows final reconstruction while the top-left one shows the image residuals. On the second row the source reconstruction (left), the potential correction (middle) and the potential correction convergence (right) are shown.}
\label{fig:potcorr}
\end{figure*}

The potential corrections are defined on a regular Cartesian grid with $21 \times 21$ pixels. Both the source and the potential have a curvature regularization (in the case of the source inversely weighted by the local image signal-to-noise ratio) and are initially over-regularized in order to keep the potential corrections in the linear regime, where the formalism of the method is valid. The potential corrections are repeated (adding the
previous correction to the current total potential), until convergence is reached in the evidence value. Results for this linear reconstruction are presented in Fig. \ref{fig:potcorr}. 

The potential correction and convergence show a clear signature of a concentrated mass over-density (i.e. a substructure) observed 
around the position $(-0.5, 1.0)''$. We have tested the effect of the potential-correction regularization on the stability of the reconstruction by using three different levels of regularization, in particular $\lambda_{\delta\psi}=10^7$, $\lambda_{\delta\psi}=10^8$ and
$\lambda_{\delta\psi}=10^9$. As expected, the convergence correction becomes smoother as the regularization increases; however the feature 
near $(-0.5, 1.0)''$ in the convergence-correction map remains clearly 
visible in each reconstruction, together with a minor mass gradient from the low right side of the ring to the up left side. This gradient is associated with the presence of the substructure itself (curvature regularization of the potential implies $\langle \kappa \rangle=0$ in the annulus and thus neither the total mass nor average convergence gradient changes in the annulus; this is an advantage of the method). For the nearly under-regularised case with $\lambda_{\delta\psi}=10^7$, the source is slightly twisted and the reconstruction becomes more noisy. This suggests that the potential allows for a minor amount of shear (we note that shear has $\kappa=0$ everywhere and that is not penalised by a curvature form of regularization), but that the substructure, although noisier, is still present
near the same position as in the other reconstruction. We therefore 
believe, given the data, that this feature is genuine. This statement, however, requires quantification. This is difficult at moment, based on the grid-based method, but can be done if the substructure is modeled 
through a simply parametrized mass component.

\subsection{Parameterized Substructure Modeling}

\begin{figure*}
    \begin{center} 
     \subfigure[]{ \includegraphics[width=0.47\hsize]{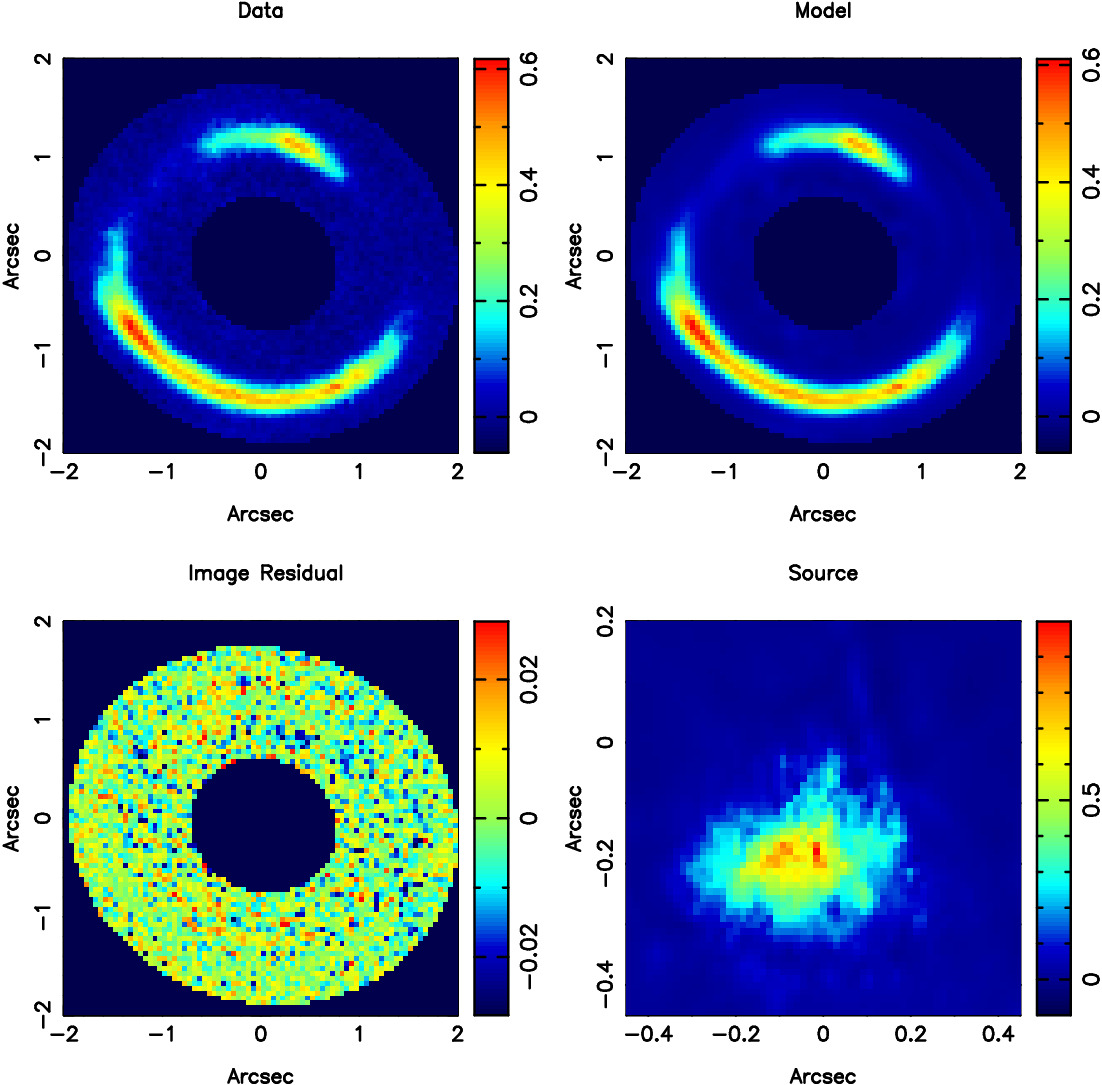}
     \label{fig:best_PL_over} 
      } 
      \hfill
      \subfigure[]{\includegraphics[width=0.47\hsize]{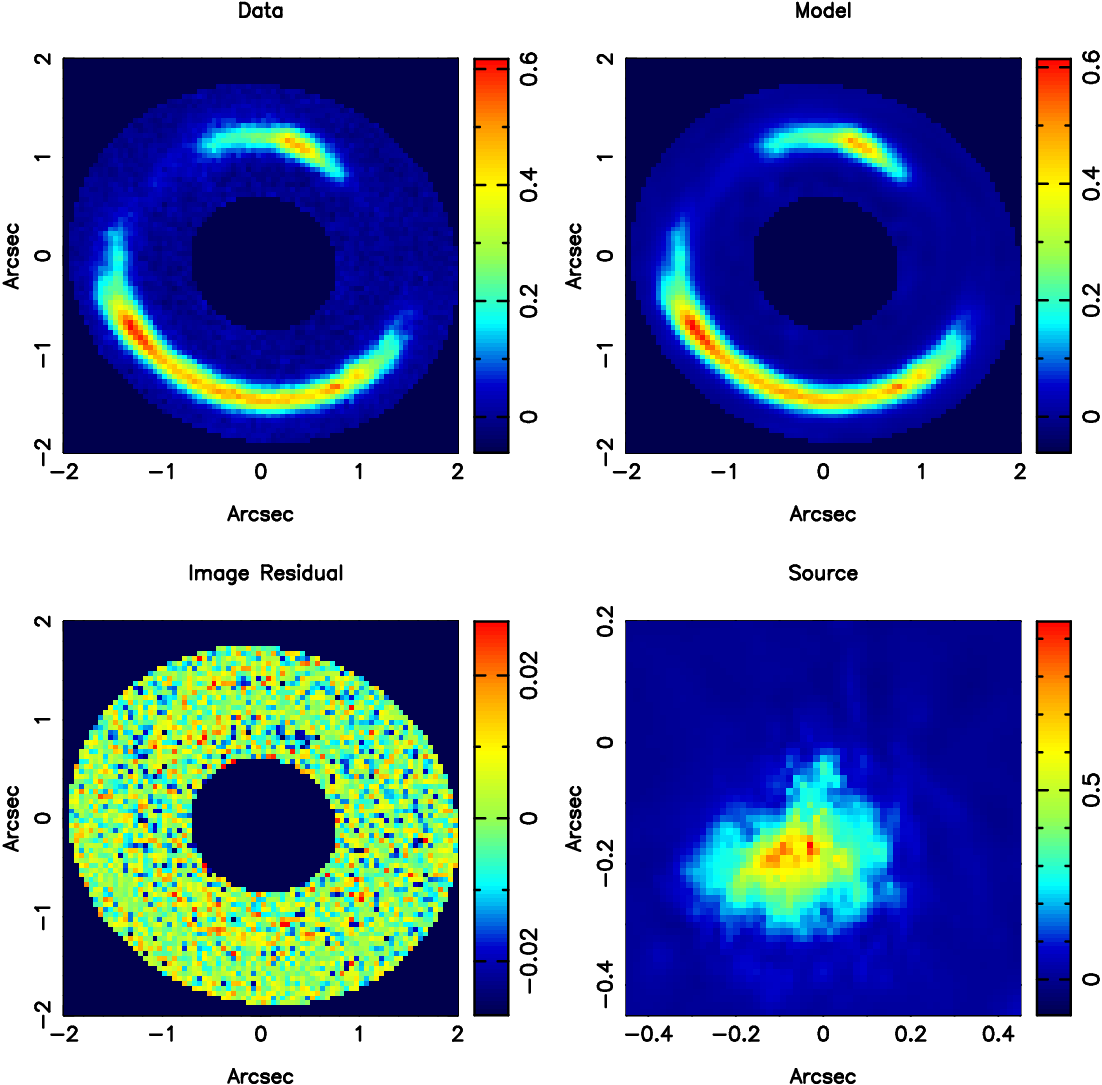}
        \label{fig:best_PJ} 
       }
      \caption{Shown are the PL+PJ model (right panel) and for the PL model with the same source regularisation as the PL+PJ model (left panel). The residuals for the PL model are still subtle but have become more pronounced and that the source model also still has more structure, despite over-regularisation. Some of these residuals are reduced by lowering the source regularisation (see text), but the evidence
difference between the two models remains large.}
    \end{center}     
 \end{figure*}

We quantify the mass of this substructure by assuming an analytic power-law (PL)+ substructure model. We assume the structure to have tidally-truncated pseudo-Jaffe (PJ) profile \citep{Dalal02} with a convergence 
\begin{equation}
k(r) = \frac{\alpha_{\rm{sub}}}{2}\left[r^{-1}-(r^2+r_t^2)^{-1/2}\right],
\end{equation}
where $r_t$ is the substructure tidal radius and $\alpha_{\rm{sub}}$ its lens strength; both are related to the main galaxy lens strength $\alpha$ and to $M_{\rm{sub}}$ by $r_t =\sqrt{\alpha_{\rm{sub}}\alpha}$ and $M_{\rm{sub}}=\pi r_t \alpha_{\rm{sub}} \Sigma_c$. 
Combining the last two relations leaves its total mass and position on the lens plane as free parameters for the substructure model. Fig. \ref{fig:best_PJ}  shows the final result of the Bayesian evidence maximisation for both the main lens and substructure parameters.

Remarkably, this procedure requires a substructure {\sl right at} the position
of the convergence overdensity found in the grid-based reconstruction. 
In terms of Likelihood, the PL+PJ model is favoured with a $|\Delta\log{\cal{L}}|=+161.0$ over the PL model (i.e.\
roughly comparable to a $\Delta \chi^{2} \sim 2\, \Delta\log{\cal{L}}$ improvement). One
might note, that the two models still seem to have similar levels of image residuals. This can be attributed to a significant difference in the source regularization. The smooth model, in order to fit the data, has to allow for more freedom to the source and has a lower level of source regularization. Hence, part of the potential structure is ``absorbed'' in the source brightness distribution. To assess the level at which the source regularization contributes to the image residual level, we run a non-linear optimisation for the smooth model while keeping fixed the regularization constant at the level of the best PL+PJ model, we call this over-regularized model $\rm{PL}_{0,\rm{over}}$ (see fig. \ref{fig:best_PL_over}). The Likelihood difference, between a perturbed model and a smooth one, is now further increased to $|\Delta\log{\cal{L}}|=+183.0$. Hence, indeed there is some covariance between the potential and source models. However, no smooth potential model does as well as models that include the PJ substructure model, near the position found in the grid-based reconstruction. We are therefore convinced, based on the Likelihood ratio, that a PL+PJ model provides a much more
probable explanation of the data, than a PL model with a more structured
source model.

Despite the difference in $\cal L({\vect{\eta}})$, at this stage of the modelling, it not possible to state whether the detection is statistically significant, because the effective number of degrees of freedom have not yet been accounted for. As shown in the next section, a nested-sampling exploration and marginalisation of the posterior probability density can be used to clarify this point and provide the Bayesian evidence values for the PL and PL+PJ models that can objectively be compared.

\section{Error analysis and Model ranking}

In this section we present the statistical analysis on the model parameters and the total marginalized evidence computation for model comparison.
We are interested to test whether the lensed images are compatible with a single smooth potential or whether the data indeed objectively require the presence of a mass substructure. We consider therefore two models, one defined by a smooth lens with a power-law density profile and one containing an additional mass substructure. In general, two models can only be objectively and quantitatively compared in terms of the total marginalized Bayesian evidence and the Bayes factor, $\Delta\log{\cal{E}} \equiv\log{\cal{E}}_{0} - \log{\cal{E}}_{1}$, which expresses their relative probability given a specific data-set. 

Heuristically the Bayesian evidence ($\cal E$) can be
compared to the classic reduced $\chi^{2}$ (i.e.\ per degree of freedom),
but without assumptions about Gaussianity of the posterior probability
distribution function about lack of covariance between parameters
(which could reduce the effective number of degrees of freedom).

\subsection{Prior Probabilities}

Prior to the data-taking, little is known about  the non-linear parameters describing the lens potential model. A natural choice is therefore a uniform prior probability.  We centre this prior on the best smooth values $\eta_{\mathrm{b,i}}$  as recovered in Section \ref{sec:smooth}, although 
the choice of prior range is not very relevant as long as the likelihood is sharply peaked inside the prior volume:
  \begin{equation}
    P\left(\eta_i\right)= \left\{
    \begin{array}{ll}
      \mathrm{constant} & \mathrm{for} \quad 
                          | \eta_{\mathrm{b,i}} - \eta_{\mathrm{i}} |
			  \leq \delta \eta_{\mathrm{i}} \\ 
        & \\ 
      0 & \mathrm{for} \quad | \eta_{\mathrm{b,i}} - \eta_{\mathrm{i}} |
          > \delta \eta_{\mathrm{i}} .
    \end{array} 
    \right.
  \end{equation} 
\noi Hence, the sizes of the intervals are taken in a such a way that they enclose the bulk of the evidence (i.e.\ likelihood times prior volume). Exactly identical priors for $\vect{\eta}$ are used for both the smooth and perturbed model. Also in the latter case, the prior is centered on the mass model parameters of the smooth model. This ensures that we are comparing their evidences in a proper manner. The regularization constant has a prior probability which is logarithmically flat in a symmetric interval around $\lambda_{s,b}$. The mass substructure is assumed to have a pseudo-Jaffe density profile and a mass with a flat prior between $M_{\rm min}=4.0\times 10^6M_\odot$ and $M_{\rm max}=4.0\times 10^9M_\odot$ \citep{Diemand07a, Diemand07b} and a position with a flat prior over the complete data grid. We note that our recovered mass, although close to the upper limit, is well inside this range (see below). We choose this range to make a comparison with simulations easier, but could have chosen a smaller or larger range. The results, however, are similar (only the evidence is offset by a constant value for both the PJ and PL+PJ models).

\subsection{The Evidence and Posterior Probability Exploration}

One of the most efficient methods for exploring the posterior probability within the framework of Bayesian statistics is the nested-sampling technique developed by \citet{Skilling04}.
Although being faster than thermodynamic integration, the nested sampling can still be computationally expensive as the overall computational cost rapidly grows with the dimension D of the problem as  $O(D^3/e^2)$, where $e$ is the desired level of accuracy \citep{Chopin07}. Most of the nested-sampling computational effort is required by the simulations of points from a prior probability distribution $\pi(\vect{\eta})$ with the constraints that the relative likelihood $\cal{L}(\vect{\eta})$ has to be larger than an increasing threshold $\cal{L}^*$.  Different approaches have been suggested in order to increase the performance of this simulation. \citet{Chopin07}, for example, proposed an extension of the nested sampling, based on the principle of importance sampling, while \citet{Mukherjee06} developed an ellipsoidal nested sampling by approximating the iso-likelihood contours by D-dimensional ellipsoids. \citet{Shaw07}, subsequently improved the ellipsoidal nested sampling with a clusters nested sampling which allows efficient sampling also of multimodal posterior distributions. 

In our analysis, we replace the standard Nested Sampling used in \citet{Vegetti09a} with MULTINEST, a multimodal nested sampling algorithm developed by \citet{Feroz08}. As further improved by \citet{Feroz09}, MULTINEST allows to efficiently and robustly sample posterior probabilities even when the distributions are multimodal or affected by pronounced degeneracies. The possibility of running the algorithm in parallel mode further reduces the computational load.

The most appealing property of nested-sampling-based techniques is that they also efficiently explore the model parameter posterior probabilities and simultaneously compute the marginalized Bayesian evidence of the model. The former provide error determinations for the parameters of a given model, while the latter allows for a quantitative and objective comparison between different and not necessarily nested (i.e. one model is not necessarily a special case of the other) models. The Bayesian evidence automatically includes the Occam's razor and penalises models which are unnecessarily complicated. This means that a PL+substructure model is preferred over a single PL only if the data require the presence of extra free parameters and the likelihood of the model increases sufficiently to offset the decrease in prior probability (i.e. extra model parameters lead to a larger prior volume and hence a smaller prior probability density near the peak of the likelihood function).   

\begin{figure*}
    \begin{center} 
     \subfigure[]{ \includegraphics[width=7cm]{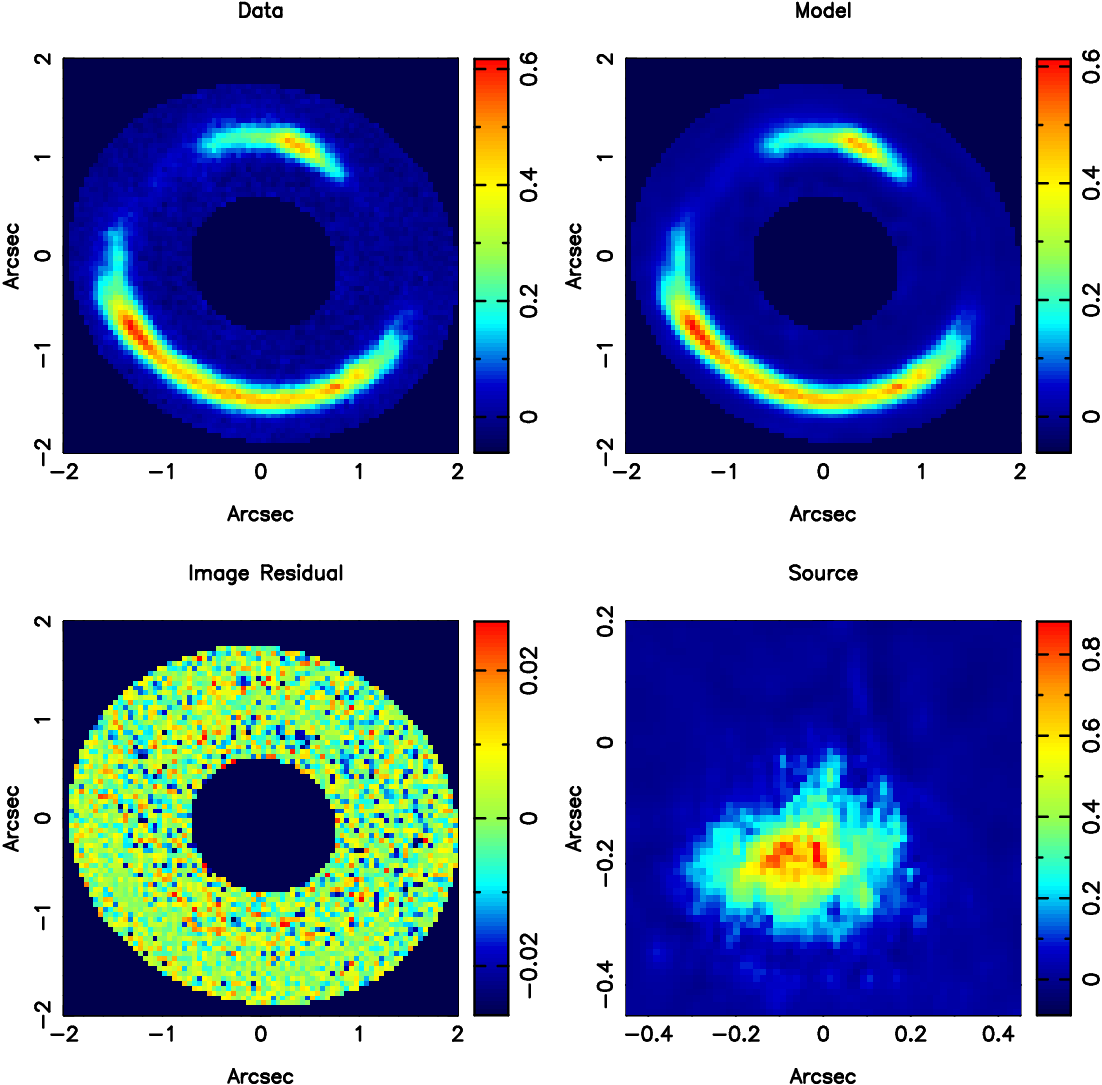}
       \label{fig:best_smooth_PSF} 
      } 
      \subfigure[]{\includegraphics[width=7cm]{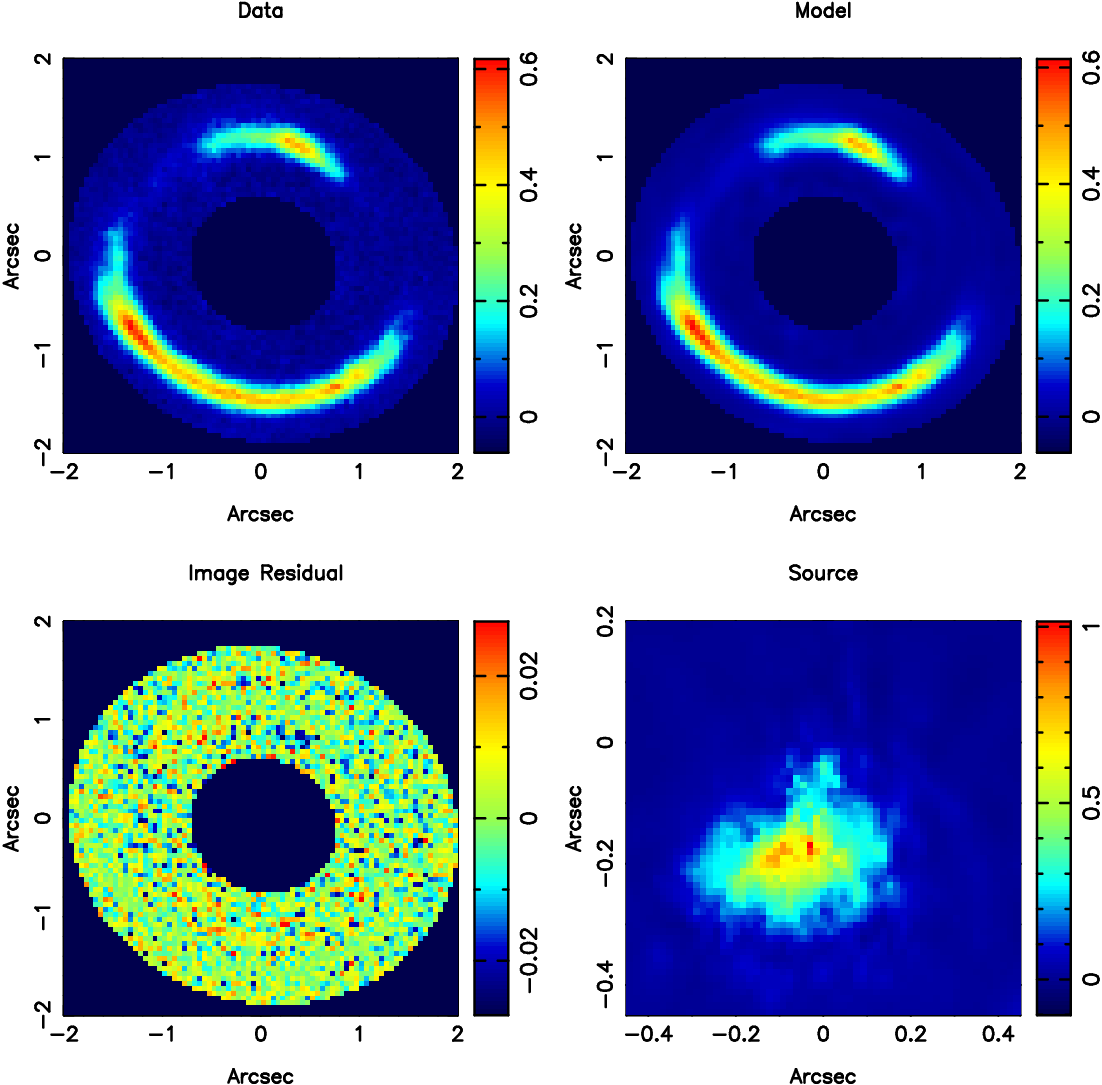}
       \label{fig:best_PJ_PSF} 
       }
      \subfigure[]{ \includegraphics[width=7cm]{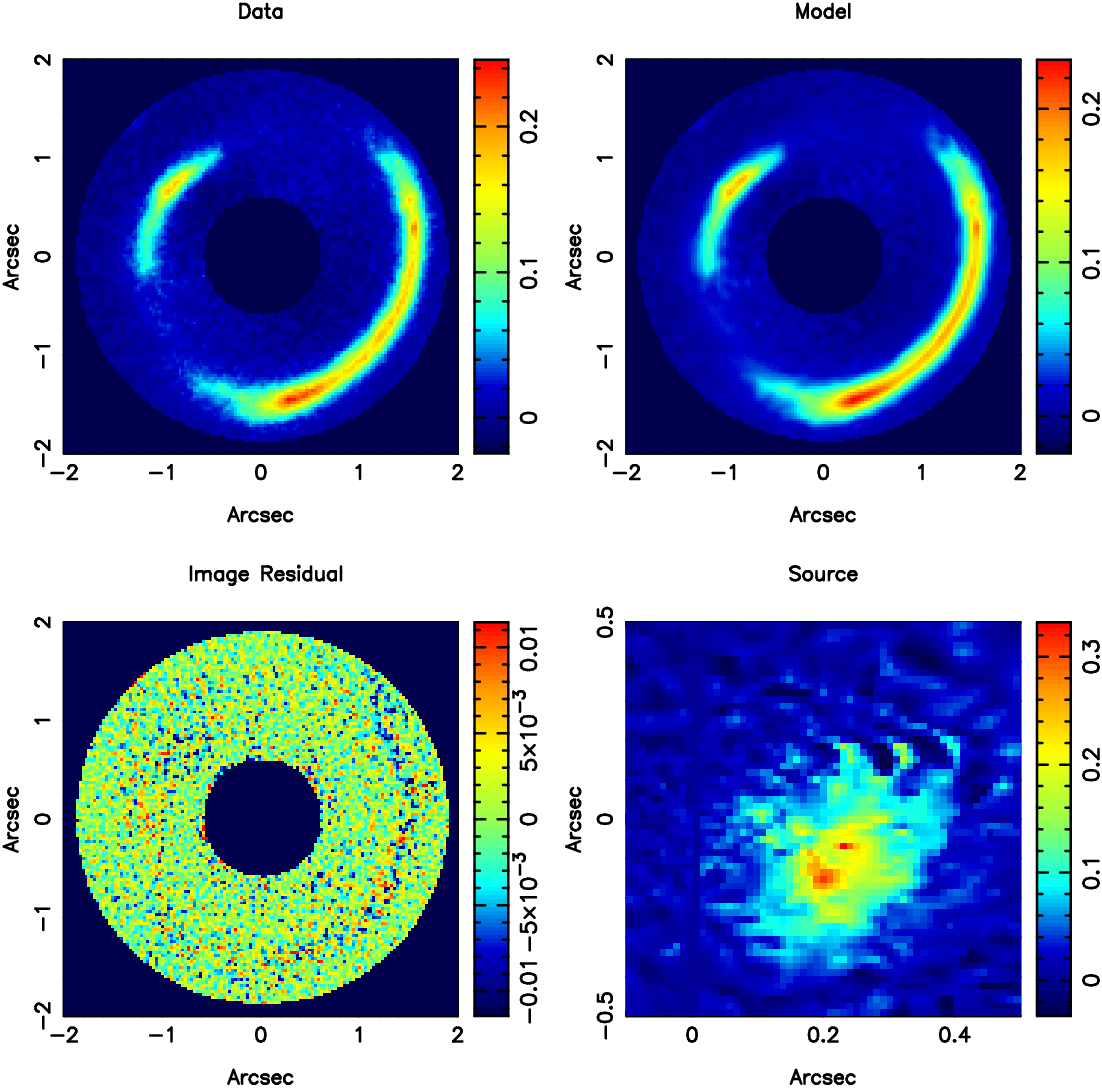}
       \label{fig:best_smooth_subt} 
      } 
      \subfigure[]{\includegraphics[width=7cm]{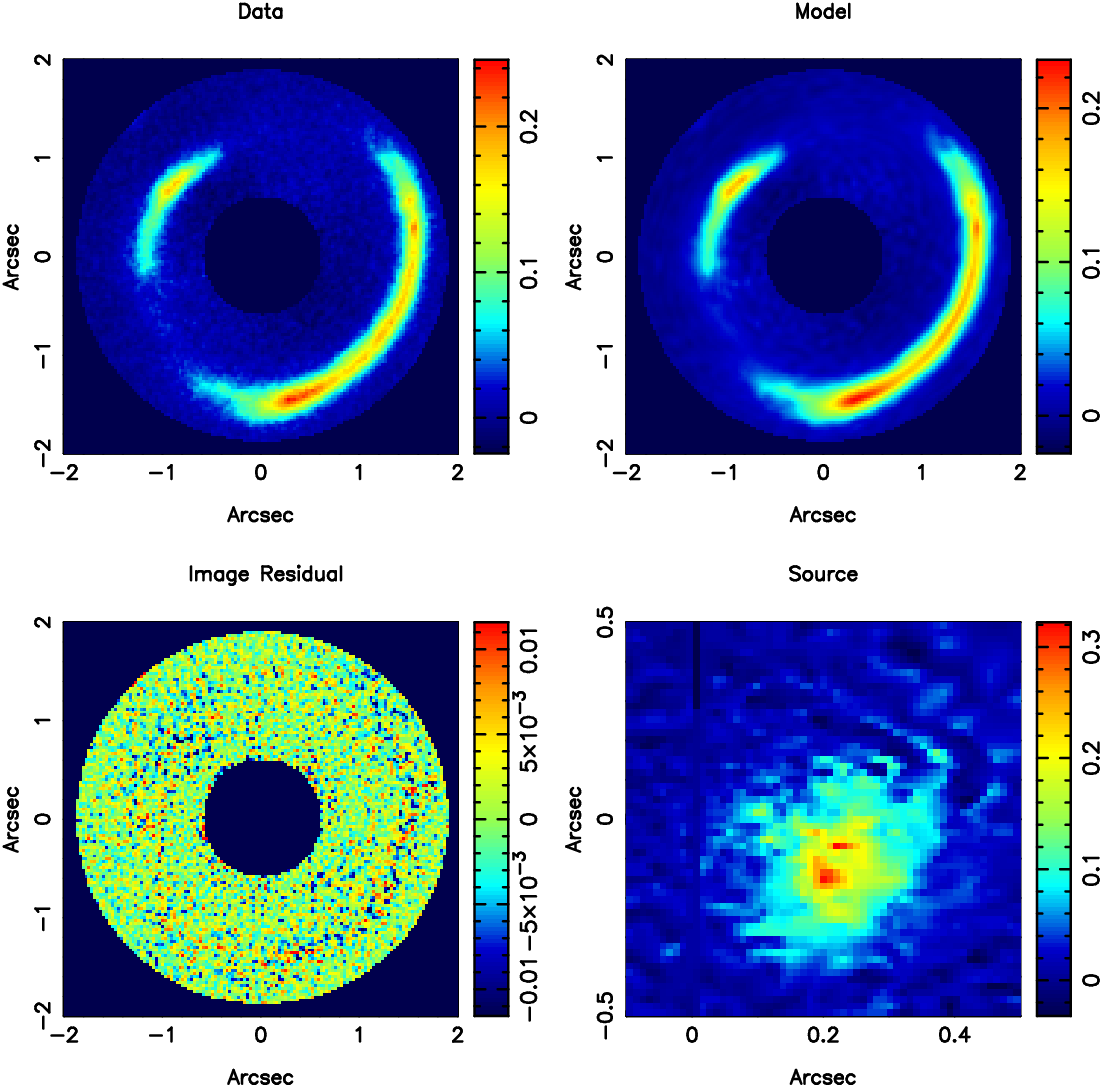}
       \label{fig:best_PJ_subt} 
       }
      \caption{{\bf Top Left panel:} The over-regularised PL model with a rotated PSF. {\bf Top Right panel:} The PL+PJ model with a rotated PSF. {\bf Bottom Left panel:} The over-regularised PL model with a smaller pixel scale and a different procedure for the lens galaxy subtraction. {\bf Bottom Right panel:} The PL+PJ model with a smaller pixel scale and a different procedure for the lens galaxy subtraction.}
        \end{center}
  \end{figure*}

\subsection{The Substructure Evidence and Model Parameters}

The main result of the Nested-Sampling analysis is that the PL+PJ model has a substructure with mean mass $$M_{\rm{sub}}=(3.51\pm 0.15)\times 10^9\msun,$$ located at a position $(-0.651\pm0.038,1.040\pm0.034)$'' (see Table 1); the quoted statistical errors do not take into account the systematic uncertainties, but fully account for all covariance in the mass model. In our case, systematic errors are mostly related to the PSF and to the procedure for  the subtraction of the lens galaxy surface brightness \citep[see][for a discussion]{Marshall07}. Effects related to systematic uncertainties are explored in Section \ref{sec:systematics}. Note that the results of this section are in agreement with those in the previous section and in particular that the substructure is exactly located where the positive convergence correction is found by the pixelized potential reconstruction (see fig. \ref{fig:potcorr}). \

Finally, we find that, the perturbed PL+PJ model is strongly favoured by the data with $\Delta\log{\cal{E}} =\log{\cal{E}}_{\rm{PJ}}-\log{\cal{E}}_{\rm{PL+PJ}}=20353.90-20482.1=-128.2$. Heuristically, and ignoring the difference
in degrees of freedom between the PL and PL+PJ models, this 
would correspond in classical terms to more or less a dramatic
$\Delta \chi^{2} \sim 256$ improvement in the model. Given that we have
thousands of data pixels and no major residuals features this shows 
that adding only a few extra parameters to the lens model improves 
the agreement between the model and the data over a wide range of 
data pixels. Heuristically one might further estimate the 
substructure mass error to be $\delta M_{\rm{sub}} \sim M_{\rm{sub}}  / \sqrt{2 {|\Delta \cal E}|} \sim 0.2 \times 10^9 \msun$, which is indeed close to the proper determination of this error.
We are therefore confident about this detection and its strong statistical significance. This represents the first gravitational imaging detection of a dark substructure in a galaxy. 

However, to test the robustness of this detection (i.e.\ systematics) we will now subject our reconstructions against several substantial changes in the model and the data, some of these going far beyond what
could be regarded as reasonable changes. 

\subsection{The Substructure Mass-to-Light Ratio}

Based on the residual images, we determine an upper limit on the magnitude
of the substructure in two different ways: by setting the limit equal to three times the estimated
(cumulative) noise level or by aperture-flux fitting, both inside a 5$\times$5 pixel (0.25$``$$\times$0.25$''$) window. The aperture is chosen to gather most of the light of  the substructure, which is expected to be effectively point-like, given the typical size of the optical counterpart of galaxies of $\approx10^9\msun$.
They are in good agreement, because the image residuals are very close to the noise level. The 3--$\sigma$
limit is found to be $I_{{\rm F814W}, 3\sigma}>27.5$~magn. At the redshift of the lens-galaxy
this corresponds to a 3--$\sigma$ upper limit in luminosity of $5.0\times 10^{6}~{\rm L}_{V,\odot}$. 
Within the inner 0.3 kpc  and 0.6 kpc,  we therefore find that the integrated mass is respectively $(5.8\pm 0.3)\times 10^8\msun$ and $(1.1\pm 0.05)\times 10^9\msun$ and hence lower limits of ${\rm (M/L)}_{{\rm V},\odot}\ga 120~ \msun/{\rm L}_{{\rm V},\odot}$ (3--$\sigma$) and ${\rm (M/L)}_{{\rm V},\odot}\ga 218 ~\msun/{\rm L}_{{\rm V},\odot}$ (3--$\sigma$). 

\begin{figure*}
    \begin{center} 
     \subfigure[]{ \includegraphics[width=0.45\hsize]{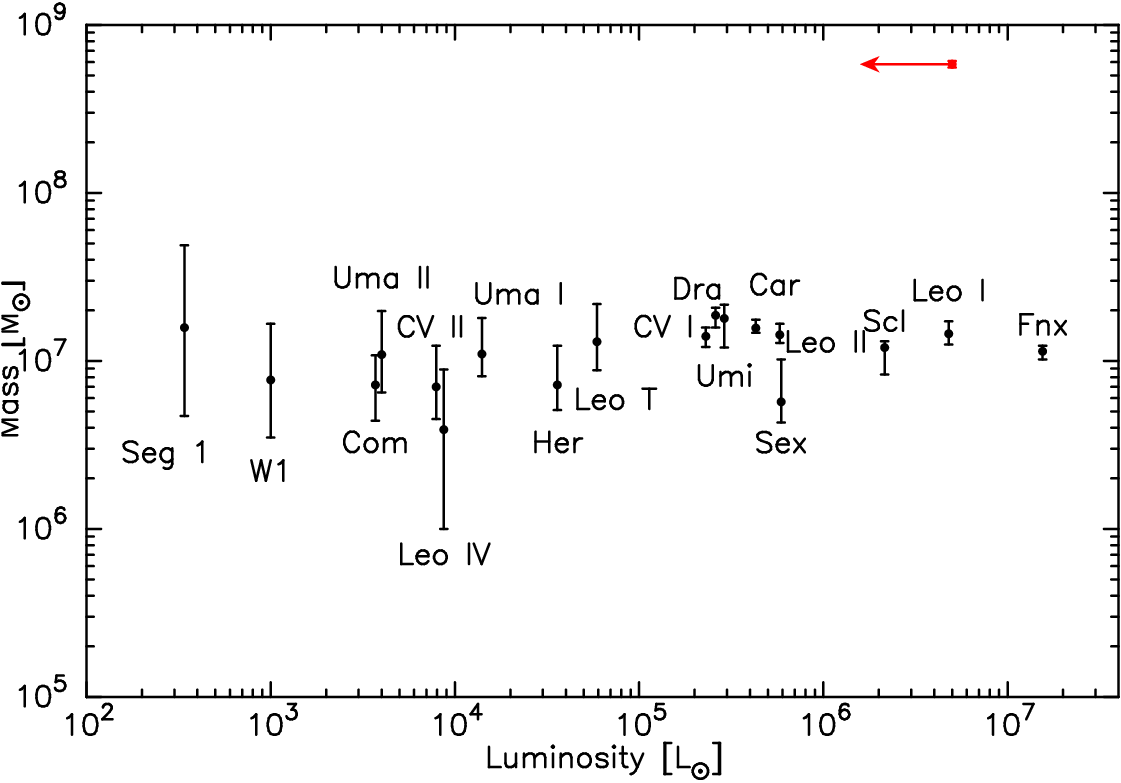}
          \label{fig:strigari} 
     	} 
      \subfigure[]{\includegraphics[width=0.45\hsize]{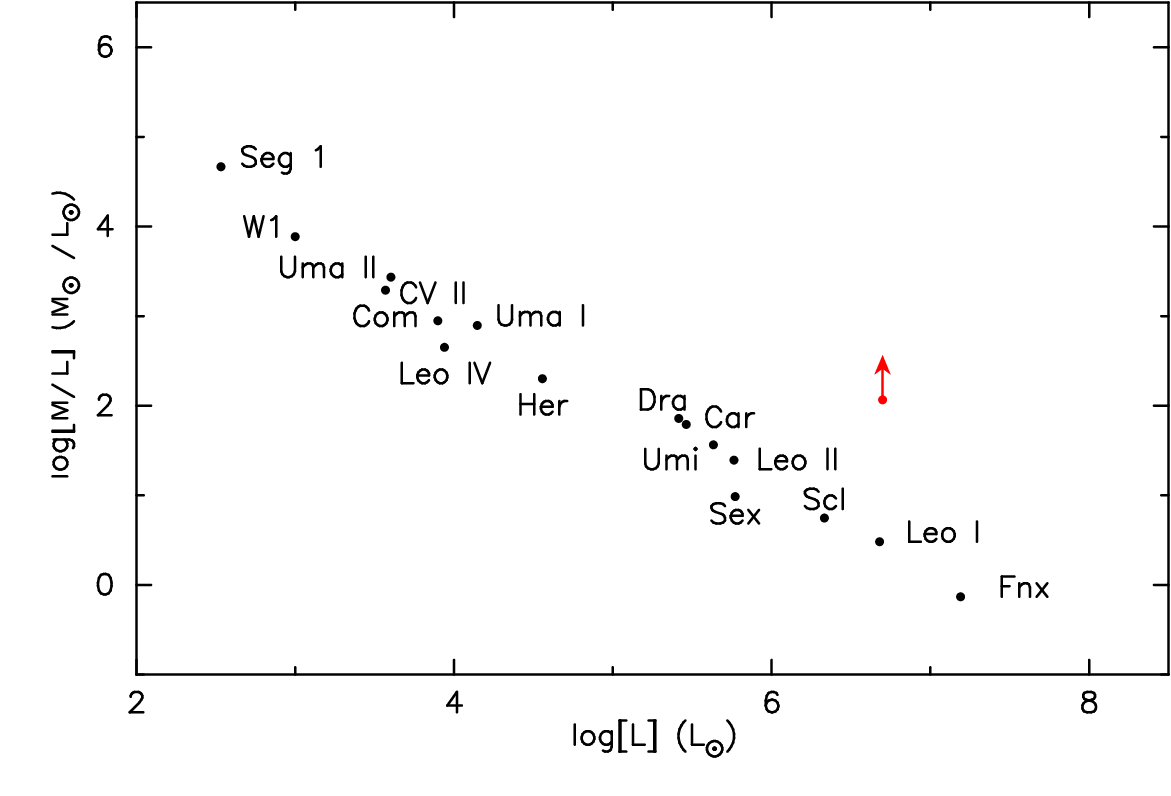}
      \label{fig:penarrubia} 
      }
      \caption{\textbf{Top panel:} The integrated mass in units of solar masses, within the inner 0.3 kpc as a function of the total luminosity, in units of solar luminosity for the Milky Way satellites (black points) and the substructure detected in this paper (red arrow). Note the small error on the substructure mass. \textbf{Bottom panel:} the integrated mass-to-light ratio, within the inner 0.3 kpc as a function of the total luminosity, in units of solar luminosity for the Milky Way satellites (black points) and the substructure detected in this paper (red arrow).}
       
\end{center}     
 \end{figure*}

The mass of the substructure is at the upper end of the mass function of Milky Way satellites (see Fig.\ref{fig:strigari} and Fig.\ref{fig:penarrubia}). This is not surprising as the normalization of the mass function scales as the total mass of the host galaxy and SDSSJ0946+1006 is substantially more massive than the Milky Way at fixed radius 
(factor $\sim$4). 
Moreover, if indeed gas is stripped from low-mass satellites though feedback and radiation in a strong star formation or starburst phase of the lens galaxy, during its formation, one might naturally expect that dwarf satellites that formed around or near massive early-type galaxies have larger M/L ratios than those in the Milky Way.
However, the total M/L upper limit is not far from those found for Milky Way satellites \citep[e.g.][]{Strigari08}
(Fig.\ref{fig:strigari}).

\begin{figure}
   \begin{center} 
      \includegraphics[width=8cm]{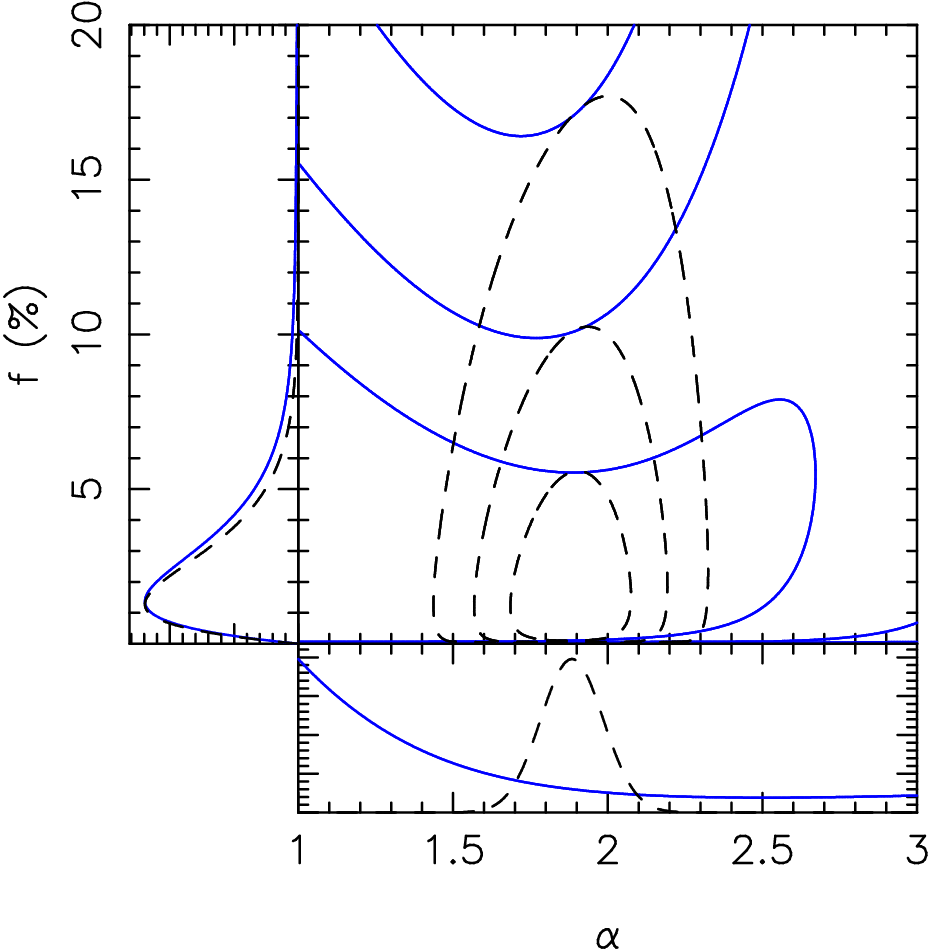}
         \caption {Joint probability $P\left( \alpha,f ~|~ \{n_s,\vect{m}\},\vect{p} \right)$ contours and marginalized probabilities 
$P\left( f ~|~\{n_s,\vect{m}\},\vect{p} \right)$ and $P\left( \alpha ~|~\{n_s,\vect{m}\},\vect{p} \right)$ for a uniform prior (solid lines) and for  a Gaussian prior in $\alpha$ (dashed lines). Contours (inside out) are set at levels $\Delta \log(P)=-1, -4, -9$ from the peak of the posterior probability density. }
      \label{fig:stats}
    \end{center}     
 \end{figure}

 \subsection{Robustness and Systematic Errors}\label{sec:systematics}
 
A number of major sources of systematic error might still affect the lens modelling: the PSF modeling, the pixel scale and lens galaxy subtraction from the lens plane. To determine at which level systematic errors influenced the substructure detection we tested the PL+PJ modelling (see Section \ref{sec:substructure}) by rotating the PSF model through $90^\circ$ from the original one; we call this model $(\rm{PL+PJ})_{\rm{psf90}}$. 
We also used a different data-set with smaller drizzled pixels (0.03$''$) and a different lens galaxy subtraction procedure (using a Sersic profile rather than a b-spline surface brightness profile); we call this model $(\rm{PL+PJ})_{\rm{subt}}$. We refer to the corresponding smooth models as $\rm{PL}_{\rm{psf90}}$ and $\rm{PL}_{\rm{subt}}$, respectively. 
The results are shown in figs. \ref{fig:best_smooth_PSF}, \ref{fig:best_PJ_PSF}, \ref{fig:best_smooth_subt} and \ref{fig:best_PJ_subt} and listed in Table 1.
 
More precisely, $(\rm{PL+PJ})_{\rm{subt}}$ is not only rotated but also has a different pixel scale (0.03''/pixel), a different number of pixels, a different noise level, and a different PSF, so that we also test against all these changes. We also check whether the form of source regularization has any effect by running a PL and a PL+PJ modelling for a non-adaptive regularization constant and for a gradient regularization. Finally we run an optimization for both the smooth and the perturbed model in which the centre of the lens is allowed to change and an optimization with a larger PSF. 

All tests (see Table 1) lead to results that are consistent with each other for both the main lens and the substructure parameters. First we note that 
rotating the PSF changes the evidence by a value that could expected 
based on the sampling error in the nested sampling. Hence we conclude
that PSF effects are negligible. In the case of the {\it subt} model, we note that we are no longer comparing the same data-sets and that the evidence
values have dramatically changed. This simply reflects the large increase 
by a factor $\sim (0.05/0.03)^{2}$ in the number of data-points. Bayesian evidence
can not be used to compare different data-set, but we can compare the
PL $_{\rm{subt}}$ and (PL+PJ)$_{\rm{subt}}$ models amongst each other. 
First, we remark that the pixel scale in this data-set is considerable smaller
than the resolution and pixel-scale in the image, hence neitherthe data
pixels nor their errors are fully independent. This leads to a rather odd stripped source reconstruction, not observed for the original data-set. 
Despite this difference, we notice that image residuals in the PL+PJ
models are reduced, especially near the substructure positions, compared
to the PL model. The Likelihood difference is $\Delta\log{\cal{L}}_{\rm{subt}} =\log{\cal{L}}_{\rm{PJ}}-\log{\cal{L}}_{\rm{PL+PJ}}=-154$ in favor of the substructure model. 

Over all, we are therefore confident that the substructure detection is not only a statistically sound detection, but also robust against dramatic changes in the model and the data. 

\section{The Substructure Mass Function}

What does this imply for the $\Lambda$CDM model and the expected 
fraction of mass in substructure? Given the statistical formalism presented in \citet{Vegetti09b}, we can use this detection to constrain the projected dark matter mass fraction in substructure $f$ and the substructure mass function slope $dN/dm\propto m^{-\alpha}$. We note that we can ignore  
the baryonic content in the substructure, because of its 
large total mass-to-light ratio.

To make a proper comparison with simulations, we assume that the substructure mass can assume any value from $M_{\rm min}=4.0\times 10^6M_\odot$ to $M_{\rm max}=4.0\times 10^9M_\odot$ \citep{Diemand07a,Diemand07b,Diemand08} and that the mass we can detect varies from $M_{\rm low}=0.15\times 10^9M_\odot$ to $M_{\rm high}=M_{\rm max}$. We note that different limits would scale both the
simulation and observed mass fraction in the same way. The
mass fractions quoted throughout the papers are for this mass range only. We ignore the error on the measured substructure mass; this does not influence the results because our detection is well beyond the error $\sigma_m=0.15\times 10^9 \msun$ level. Given the quality of the fit for a PL+PJ model, we are confident that there are no other substructures with mass above our detection threshold.

Figure~\ref{fig:stats} shows the joint posterior probability density function $P\left(\alpha,f \,|\, n_s=1,m=M_{\rm{sub}},\vect{p} \right)$ contours and the marginalized probability densities $P\left(f\,|\,n_s=1,m=M_{\rm{sub}},\vect{p} \right)$ and $P\left( \alpha ~|~n_s=1,m=M_{\rm{sub}},\vect{p} \right)$, given one detected substructure $n_s=1$ with mass $m=M_{\rm{sub}}$; where $\vect{p}$ is a vector containing the model parameters,  $M_{\rm min}$,  $M_{\rm max}$,  $M_{\rm low}$ and  $M_{\rm high}$. Specifically, from the mariginalized probability density distributions we find $f=2.56_{-1.50}^{+3.26}\%$ and $\alpha=1.36_{-0.28}^{+0.81}$ at a 68\% confidence level for a flat prior on $\alpha$ and $f=2.15_{-1.25}^{+2.05}\%$ and $\alpha=1.88_{-0.10}^{+0.10}$ at a 68\% confidence level for a Gaussian prior centred in $1.90\pm0.1$. The same results are found if an error on the mass measurement and a detection threshold $M_{\rm low}=3 \times \sigma_m$ are assumed. 

\noi As already discussed in \citet{Vegetti09b}, while even a single lens system is enough to set upper and lower limits on the mass fraction, a larger number of lenses is required in order to constrain the mass function slope, unless a stringent prior information on the parameter it-self is adopted. By assuming a Gaussian prior of the mass function slope centred in 1.90, we can quantify the probability that the dark matter mass fraction is the one given by N-body simulations $f \approx 0.3\%$ \citep{Diemand07a,Diemand07b,Diemand08}, by considering ratio between the posterior probability densities  $P\left( f _{\rm{N-body}}, \alpha=1.9|~n_s=1,m=M_{\rm{sub}},\vect{p} \right)$  and $P_{\rm{max}}\left( f, \alpha=1.9 \, |~n_s=1,m=M_{\rm{sub}},\vect{p} \right)$ and find that this
ratio is 0.51. Hence, currently our measurement and that inferred from N-body simulations
still agree as a result of the rather larger error-bar on the measured value of $f$. The combination of more lens systems is, of course, required to set more stringent constraints also on $\alpha$. We plan such an analysis in forthcoming papers. 

\noi Given our best value of $f=0.0215$ for $\alpha=1.9$, we might 
expect to detect $\sim$1 mass substructures above our 3--$\sigma$ mass-threshold of $4.5\times 10^{8}\msun$. It is therefore unlikely that we 
have missed many substructures with a mass slightly below that of our detection. Given this result and image residuals already at the 
noise level, we believe that adding a second substructure is not warranted
and missing lower-mass substructure leads only to a minor bias (note
that logarithmic bins have nearly equal amounts of mass for $\alpha=1.9$).

 \begin{table*}
    \begin{center}
      \caption {Parameters of the mass model distribution for the lens SDSSJ0946+1006. For each parameter we report the best recovered value and the relative Likelihood for a smooth model (PL) in column (2), for a smooth over-regularized smooth model in column (3), for a perturbed model (PL+PJ) in column (4), for a smooth and perturbed model (PL+PJ) with rotated PSF respectively in columns (5) and (6) and a smooth and perturbed model (PL+PJ) for different galaxy subtraction respectively in columns (7) and (8).
      We note that the models in the final two columns use a different (also rotated) data set
      and the evidence values, position angles and positions can therefore not be directly compared.}
      \begin{tabular}{ccccccccc} 
	\hline
	&$(\rm{PL})_0$&$\rm{PL}_{0,\rm{over}}$&$(\rm{PL+PJ})_0$&$\rm{PL_{\rm{psf90}}}$&$(\rm{PL+PJ})_{\rm{psf90}}$&$\rm{PL}_{\rm{subt}}$&$(\rm{PL+PJ})_{\rm{subt}}$\\
	\hline
	$\alpha~({\rm arcsec})$&1.329&1.329&1.328&1.329&1.328&1.280&1.272\\
	\\
	$\theta~({\rm deg})$&65.95&65.80&69.26&64.97&71.04&-60.99&-60.96\\
	\\
	$f$&0.961&0.961&0.962&0.962&0.963& 0.982&0.982\\
	\\
	$q$&0.598&0.597&0.599&0.597&0.600&0.641&0.646\\
	\\
	$\Gamma_{sh}~({\rm arcsec})$&0.081&0.081&0.086&0.080&0.087&-0.092&-0.097\\
	\\
	$\theta_{sh}~({\rm deg})$&-20.83&-20.65&-22.32&-20.63&-22.12&-39.83&-40.58\\
	\\
	$\log(\lambda_s)$&1.152&2.028&2.028&1.059&1.988&0.036&0.052\\
	\hline
	 $m_{\rm{sub}}~(10^{10}\msun)$&&&0.323&&0.333&&0.342\\
	\\
	$x_{\rm{sub}}~({\rm arcsec})$&&&-0.686&&-0.682&&-1.286\\
	\\
	$y_{\rm{sub}}~({\rm arcsec})$&&&0.989&&0.9956&&-0.391\\
	\\
	   \hline
	$\log\cal{L}$&20350.97&20328.11&20511.14&20358.49&20525.32&61520.63&61674.63\\
	\hline
      \end{tabular} 
      \label{tab:results}
    \end{center}
  \end{table*}

\section{Summary}

We have applied our new Bayesian and adaptive-grid method for pixelized source {\sl and} lens-potential modeling \citep{Vegetti09a} to the analysis of HST data of the double Einstein ring system SLACS SDSSJ0946+1006 \citep{Gavazzi08}. This system was chosen based on its large expected dark-matter mass fraction near the Einstein radius and the high signal-to-noise ratio of the lensed images.
Although these two facts should be uncorrelated to the mass fraction of CDM substructure, both
incidences maximize the change of detection.

We find that a smooth elliptical power-law model of the system leaves significant residuals near or above the noise level; these residuals are correlated and spread over a significant part of the lensed images. Through a careful modeling of this data including either lens-potential corrections or an additional (low-mass) simply parametrized mass component, we conclude that the massive early-type lens galaxy of SLACS SDSSJ0946+1006 hosts a large mass-to-light ratio substructure with a mass around $M_{\rm{sub}}\sim 3.5 \times 10^9\msun$, situated on one of the lensed images. 
A careful statistical analysis of the image residuals, as well as a number of more drastic robustness tests (e.g.\ changing the PSF, pixel number and scale, regularization level and form, galaxy subtraction and image rotation), confirm and support this detection. 
Based on this detection, the first of its kind, we derive a projected CDM substructure mass fraction of $\sim 2.2\%$ for the inner regions of the galaxy, using the Bayesian method of \citet{Vegetti09b}; this fraction is high, but still consistent with expectations
from numerical simulations due to the large (Poisson) error based on a single detection. 

The numerical details of our results can be further summarized as follows:

\begin{enumerate}

\item[{\bf (1)}] Using a Bayesian {\tt Multinest} Markov-Chain exploration of the full model parameter space, we show that the identified object has a mass of $M_{\rm{sub}}=\left(3.51\pm0.15\right)\times 10^9\msun$  (68\% C.L.) and is located near the inner Einstein ring at $(-0.651\pm0.038,1.040\pm0.034)''$. The Bayesian evidence is in
favor a model that includes a substructure versus a smooth elliptical
power-law only, by $\Delta \log({\cal E})=-128.2$. This is roughly equivalent
to a $16-\sigma$ detection. 

\medskip

\item[{\bf (2)}] At the redshift of the lens-galaxy
a 3--$\sigma$ upper limit in luminosity is found of $5.0\times 10^{6}~{\rm L}_{V,\odot}$. Within the inner 0.3 kpc  and 0.6 kpc,  we find that the integrated mass is respectively $(5.8\pm 0.3)\times 10^8\msun$ and $(1.1\pm 0.05)\times 10^9\msun$ and hence lower limits of ${\rm (M/L)}_{{\rm V},\odot}\ga 120~ \msun/{\rm L}_{{\rm V},\odot}$ (3--$\sigma$) and ${\rm (M/L)}_{{\rm V},\odot}\ga 218 ~\msun/{\rm L}_{{\rm V},\odot}$ (3--$\sigma$). This is higher than of MW
satellites, but maybe not unexpected for satellites near massive elliptical galaxies.

\medskip

\item[{\bf (3)}]  The CDM mass fraction and a mass function slope are equal to  $f=2.15_{-1.25}^{+2.05}\%$ and $\alpha=1.88_{-0.10}^{+0.10}$, respectively,  at a 68\% confidence level for a Gaussian prior on $\alpha$ centred on $1.90\pm0.1$. For a flat prior on $\alpha$ between 1.0 and 3.0, we find $f=2.56_{-1.50}^{+3.26}\%$. Asking whether the $f=2.15\%$
is consistent with $f=0.3\%$, we find a likelihood ratio of 0.51; indeed both are consistent. This is 
the result of the considerable measurement error found for $f$, because it is based on only a single 
detection. 

\end{enumerate}

\noi This is the first application of our adaptive ``gravitational imaging'' method to real data and clearly shows its promise. In the near future we will apply the method to a larger set of SLACS lenses in order to constrain, via the statistical formalism presented in \citet{Vegetti09b}, the dark matter fraction in substructure and the substructure mass function.
 
\vfill 
 
 \section*{Acknowledgements} 
S.V. is grateful to Farhan Feroz for the help provided in the implementation of MULTINEST. We thank Simon White and Chris Kochanek for useful suggestions.

S.V. and L.V.E.K. are supported (in part) through an NWO-VIDI program subsidy (project number 639.042.505). TT acknowledges support from the NSF through CAREER award NSF-0642621, by the Sloan Foundation through a Sloan Research Fellowship, and by the Packard Foundation through a Packard Fellowship. 
This research is supported by NASA through Hubble Space Telescope programs SNAP-10174, GO-10494, SNAP-10587, GO-10798, and GO-10886.
Based on observations made with the NASA/ESA Hubble Space Telescope, obtained at the Space Telescope Science Institute, which is operated by the Association of Universities forResearch inAstronomy, Inc., under NASA contract NAS 5-26555. Funding for the creation and distribution of the SDSS Archive has been provided by the Alfred P. Sloan Foundation, the Participating Institutions, theNationalAeronautics and Space Administration, the National Science Foundation, the US Department of Energy, the Japanese Monbukagakusho, and the Max Planck Society. The SDSS Web site is http://www.sdss.org. The SDSS is managed by the Astrophysical Research Consortium (ARC) for the Participating Institutions. The Participating Institutions are the University of Chicago, Fermilab, the Institute for Advanced Study, the Japan Participation Group, Johns Hopkins University, the Korean Scientist Group, Los Alamos National Laboratory, the Max-Planck-Institute for Astronomy (MPIA), the Max-Planck-Institute forAstrophysics (MPA),NewMexico State University, University of Pittsburgh, University of Portsmouth, Princeton University, the United States Naval Observatory, and the University of Washington.


\end{document}